%% file: RJwrapper.tex
\documentclass[a4paper]{report}
\usepackage[utf8]{inputenc}
\usepackage[T1]{fontenc}
\usepackage{RJournal}
\usepackage{amsmath,amssymb,array}
\usepackage{booktabs}

\begin{document}

\sectionhead{ }
\volume{ }
\volnumber{ }
\year{ }
\month{ }

\begin{article}
  \input{scarpino-gillete-crews}

\end{article}

\end{document}

%% file: scarpino-gillete-crews.tex
\title{\pkg{multiDimBio}: An R Package for the Design, Analysis, and Visualization of Systems Biology Experiments}
\author{Samuel V. Scarpino, Ross Gillette, David Crews}

\maketitle

\abstract{
The past decade has witnessed a dramatic increase in the size and scope of biological and behavioral experiments.  These experiments are providing an unprecedented level of detail and depth of data.  
However, this increase in data presents substantial statistical and graphical hurdles to overcome, namely how to distinguish signal from noise and how to visualize multidimensional results.  Here we present a 
series of tools designed to support a research project from inception to publication.  We provide implementation of dimension reduction techniques and visualizations that function well with the types of data often 
seen in animal behavior studies.  This package is designed to be used with experimental data but can also be used for experimental design and sample justification.  The goal for this project is to create a package that will evolve over time, thereby remaining relevant and reflective of current methods  and techniques.
}

\section{Introduction}
Two terms, trait and phenotype, are fundamental in biology. Traditionally, a trait is defined as any character of an organism that can be described and/or measured, while a 
phenotype is the sum total of the measured traits. For example, in systematics, traits are used as diagnostics, and phenotypes as descriptors, of species; in other words, traits are the divisible 
units of phenotypes and phenotypes are assemblages of traits. Although traits originally referred to observable characters such as morphology and behavior, this definition was expanded with 
technical advances to include the physiological, developmental, and genetic processes that produce these characters (~\cite{Moore1997, Nijhout2005}). Despite these important distinctions, it is not uncommon 
to see these two terms used synonymously. For example, in studies of genetically modified organisms one often sees ``phenotype" referring to the product of the gene manipulation while other traits, that may or 
may not have been measured, are disregarded.

Another issue to keep in mind is that all traits, whether it is a morphological feature of the organism, its behavior, physiology, or even the patterns of gene expression during the formation of a 
tissue, change throughout the life history of the individual, which lead to changes in the phenotype (~\cite{Crews2006b}).  Even the genome might change under unusual circumstances such as environmentally 
induced mutation (~\cite{Crews2006}). This history can be evaluated on a scale of seconds, minutes, days, an individual's lifetime, or at the population level over the course of generations. The challenge then is 
how to calculate this constant yet ever changing complexity in ways that inform, rather than oversimplify.   

Many classic studies have focused on the outcome of manipulating a single variable in an experimental setting, e.g. vaccine exposure prior to challenge with an infectious agent.  However, 
these studies oversimplify the phenotype, relegating the phenotype to a simple trait as opposed to the manifestation of a complex network of traits.  A systems biology approach can place the 
results of experiments in the broader context of how phenotypes arise from sets of interacting traits (~\cite{Kitano2002}).  These analyses can be restricted to single levels of biological 
organization or can be used to integrate across levels; either alternative gives a fuller understanding of the phenotype of the individual. Fortunately, more researchers are coming to appreciate 
the interrelated nature of traits at all levels of biological organization and, in so doing, have begun to look at the interactions of traits rather than considering them as if they were independent 
variables.

When multiple traits of complex phenotypes are examined as a unit, $($e.g., the suite of genes known to be involved in sex determination and gonadal differentiation or the neural circuitry 
underlying sociosexual behavior$)$, conventional analytic and presentation methods make it difficult to quantify and illustrate the information. This paper introduces an adaptation of 
established methods for analyzing complex data sets that takes a computational systems biology approach, integrating data, analysis, and visualization. Our method of depicting complex 
phenotype analysis, which we have called the Functional Landscape Method, can be viewed as a recent addition to the long history of imagery to depict complex concepts in all areas of 
science. Well-known images in Biology would include the~\cite{Waddington1957} developmental landscape depicting the genes that shape tissues and, more recently, the~\cite{Nijhout2003}
schematic of the importance of context in trait development. Similarly, in Psychology, there is the ~\cite{Gottesman1997} depiction of the contribution of genes to cognitive ability and that of ~\cite
{Grossman2003} illustrating how genetic and experiential factors push the individual to thresholds of pathology.  Notably, all share the use of three dimensions to illustrate complex traits whose 
individual components are two-dimensional in nature. The shared quality of these images is predicated on the fact that the mind can process 3D comparisons much better than complex bar 
graphs or tabulated results, a fact verified many times in cognitive psychology. 

\section[Models Implemented in multiDimBio]{Models Implemented in \pkg{multiDimBio}}
\pkg{multiDimBio} was designed to support a research project involving multidimensional data from design through publication and as such the package will continue to grow and change as new methods are 
developed.  Researchers interested in contributing to \pkg{multiDimBio} are encouraged to contact the authors.  The models implemented section is organized from design to analysis and visualization, as one 
would conduct a research project.

\subsection{Power Analysis}
Methods are implemented to compute the statistical power, in terms of the type II error rate, based on anticipated sample and effect sizes for \code{FSelect()} and \code{PermuteLDA()}.  By default, the power of 
both tests are determined by iterating over a range of effect and sample sizes.  The default settings were selected to be representative of many behavioral genetic studies; however, users can input alternative 
sample and effect sizes.  The algorithm for the power analysis proceeds as follows:

\begin{enumerate}
\item  Input sample and effect sizes\item  Set the number of significant effects, e to 0. \emph{Note - } Total number of traits is fixed at 6\item  Draw random deviates for the given sample size for 6 traits. \emph{Note 
- } All traits not significant under this iteration are drawn from a N(0,1) distribution.\item   Perform either \code{FSelect()} or \code{PermuteLDA()} and record the results.\item  Return to step 3 N times, recording 
the results each time. \emph{Note - } N is set using the trials input\item  If e<5 return to step 2 and set the number of significant effects to e+1\item  Proceed to the next combination of sample and effect size.\item  
Output the results for each combination of sample and effect size as a function of the number of significant traits.
\end{enumerate}

\code{Power(func, N, effect.size, trials)}
\begin{itemize}
\item \code{func}  = The function being used in the power analysis, either PermuteLDA or FSelect.\item \code{N}= A vector of group sizes.  The group size is N/2. Default = c(0.1,0.4,0.8,1.6)\item \code{effect.size} 
= A vector of effect sizes.  Default = c(6,12,24,48,96)\item \code{trials} = The number of iterations for each combination to determine the type II error rate.
\end{itemize}

\subsection{Data Preprocessing} \label{PP}
A number of methods are implemented in \pkg{multiDimBio} to aid in the necessary preprocessing of data. 
\paragraph{Incomplete data} \label{ID}
A recurrent challenge in analyzing data from behavioral research is missing observations.  Because our methodologies require individuals with the same number of observations, addressing 
this problem is critical.  We use a three-step process to solve the missing data problem.  First, all traits measured on fewer than 50 percent of the individuals are removed.  Second, all
individuals missing more than 50 percent of the remaining traits are removed.  Importantly, the user can modify each of these arbitrary thresholds.  Finally, missing data is imputed using a probabilistic principle component framework.  Our implementation is a 
wrapper around the \BIOpkg{pcaMethods} functions \code{ppca} and \code{svdimpute} (~\cite{Stacklies2007}).  Unlike traditional 
principle component analysis, probabilistic principle component analysis (PPCA) can handle missing data (~\cite{Tipping1999}).  In the implementation of PPCA  used in \pkg{pcaMethods} an Expectation Maximization (EM) algorithm is used 
to fit a Gaussian latent variable model (~\cite{Tipping1999}).  Missing values are then imputed as a linear combination of the principle components (~\cite{Troyanskaya2001}).  The output is a complete matrix of principle component scores, a vector of 
individuals and/or traits removed during the first two steps, and a diagnostic plot illustrating the performance of the data estimation step.  Although not currently implemented, one could also estimate the statistical uncertainty associated with each interpolated point.   We use the following algorithm to assess the 
performance of the data imputation:

\begin{enumerate}
\item Determine the percentage of missing data from the complete data frame, {\it pmiss}.\item Remove all individuals without complete observations.\item Randomly censor the observations with probability  {\it 
pmiss}.\item Use the three-step method outlined in section~\ref{ID} to impute the missing data.\item Compare the observed data to the imputed.\item Analyze the residuals using a \code{qq-plot}.  If the method 
performs appropriately the error should be normally distributed. 
\end{enumerate}

The PPCA method was adapted from ~\cite{Roweis1997} and a Matlab script developed by Jakob Verbeek. The data estimation method was proposed by ~\cite{Troyanskaya2001}. 

\code{CompleteData(DATA, cut.trait=0.5, cut.ind=0.5,show.test=TRUE)}
\begin{itemize}
\item \code{DATA}  = a non-empty numeric matrix with missing values\item \code{cut.trait} = Threshold for removing a trait, 0 = remove all traits, 1 = keep all traits\item \code{cut.ind} = Threshold for removing an 
individual, 0 = remove all individuals, 1 = keep all individuals \item \code{show.test} = Logical, TRUE = run a diagnostic on the methods performance and plot the results, FALSE = do not run a diagnostic.
\end{itemize}

\paragraph{Centering and scaling methods}
Multivariate analyses are often sensitive to traits with significantly different means or variances from other traits and therefore it is necessary to center and scale the data.  There have been 
countless ways proposed to center and scale data, so please consider these only as suggested methods.  \code{PercentMax()} scales all data by the maximum value observed for that trait, 
as a result each trait score will have a range from 0 to 1, with 1 indicating the maximum observed value.  \code{ZTrans()} converts each trait measurement into a z-score based on the mean 
and variance of the trait.  \code{MeanCent()} centers each trait to have mean 0.  \code{MeanCent()} can be combined with one of several proposed methods for scaling the variance of 
a population

\subsection{Multivariate Analyses}
The multivariate analyses provided in this package are not novel.  It is their joint application to data that offers a new approach to analyzing higher-dimensional data.  Multivariate Analysis of 
Variance, MANOVA, has been an established statistical approach for analyzing multivariate data for decades.  The underlying statistical framework for \code{FSelect} was developed in the 
1970s and 80s (see ~\cite{Jennrich1977} for step-wise discriminant analysis and ~\cite{Costanza1979} for the use of F-statistics).  \code{PermuteLDA} was developed by ~\cite
{CollyerAdams2007} as a method to perform linear discriminant analyses on sparse data sets.

\paragraph{MANOVA}
MANOVA is performed using the  R function \code{manova} from the \CRANpkg{stats} package and is the classic implementation of the method in R (~\cite{stats}).  Readers unfamiliar with this 
method should see ~\cite{Everitt2011}.  The MANOVA functionality of this package was designed in combination with interaction plots and categorical variables to control for random effects.  

\paragraph[Fselect]{\code{FSelect}}
~\cite{Habbema1977} first described the use of F-statistics to select variables for inclusion in a discriminant analysis. The rational was that selecting variables for inclusion in a discriminant analysis should be 
based on criteria associated with the desired use of the test, in this case to differentiate two or more groups.  F-statistics are an ideal metric in this case because they summarize how different two groups are. In 
the current implementation of the method only two groups can be compared; this arose from the complication of calculating partial F-statistics when there are more than two groups.  Partial F-statistics are 
calculated using equation~\ref{eq:partF}.  

\begin{equation} \label{eq:partF}
F_{partial} = (v - p +1)*\frac{F_{full} - F_{1}}{v + F_{1}},
\end{equation}
 
where $n1 = $ number of individuals in group 1, $n2 = $ number of individuals in group 2, $p = $ the number of discriminant axes, $F_{full} = $ the F statistic for the full model, $F_{1} = $ the F statistic for the 
single trait, and $v = n1 + n2 - 2$. \code{FSelect} performs discriminant analysis using the \code{lda} function as implemented in the \CRANpkg{MASS} package and corrects for multiple comparison using the \code
{p.adjust} function in the \CRANpkg{stats} package (~\cite{MASS, stats}).

The algorithm proceeds as follows:
\begin{enumerate}
\item Input the group IDs, data, and desired number of axes\item Set the number of selected axes, s, equal to 0 and the number of non-selected axes, c, equal to the number of columns imputed.\item For each 
axis in c perform a linear discriminant analysis using only that trait and calculate the resulting F-statistic.\item Select the axis, trait, with the largest F-statistic and move that axis from c into s.\item Set s = 1\item For 
all axes in c perform a linear discriminant analysis using that axis and all axes in s.  \item Calculate the partial F-static for each model using equation~\ref{eq:partF}.\item Select the trait with the largest partial F-
statistic and move that trait from c into s.\item Set s to s+1\item If s is < the desired number of axes and c>0 return to step 6.\item Perform a final model using all axes in s and calculate a p-value for each axis.\item 
Control for multiple comparisons using a pre-specified method, the default being a Bonferroni-Holm correction as implemented in the stats package function p.adjust\item Output the results.
\end{enumerate}

\code{FSelect(Data,Group, target, p.adj.method="holm", Missing.Data="Complete")}
\begin{itemize}
\item \code{Data}  = a (non-empty) numeric matrix of data values\item \code{Group} = a (non-empty) vector indicating which group the rows in Data belong to\item \code{target} = Number of axes selected \item 
\code{p.adj.method} = Multiple comparison correction.  Using the function \code{p.adjust} in \pkg{stats}\item \code{Missing.Data} = Missing data must either be removed or imputed.  Complete uses \code
{CompleteData} to impute missing data, Remove censors individuals with missing values for one or more traits.
\end{itemize}

\paragraph[PermuteLDA]{\code{PermuteLDA}}
Determining the statistical significance of a discriminant function analysis along with performing that analysis on sparse data sets, e.g., many traits observed on comparatively few individuals, is a challenge. ~
\cite{CollyerAdams2007} developed a Monte Carlo based algorithm for addressing both of those issues.  Briefly, the test uses the underlying Var$/$Cov structure of the data and randomizes the group 
membership to calculate a null distribution.  This test simultaneously controls for heteroscedasticity, a common problem in sparse data sets, and allows the approximation of a p-value for the test.  For the original 
implementation and formulation of the method see~\cite{CollyerAdams2007}  or \url{http://www.public.iastate.edu/~dcadams/software.html}.  Unlike the \code{FSelect} implementation, \code{PermuteLDA} will 
work properly with an arbitrary number of groups.  The time required to run the algorithm is non-linear in the number of groups.  The algorithm proceeds as follows:
\begin{enumerate}
\item Input the data and group IDs.\item A pairwise comparison matrix is generated such that each possible unique pair of groups will be analyzed.  The number of combinations is simply choose(n,2), where n is 
the total number of groups.\item Fit a linear model with group ID as the independent variable and the data as the dependent variables.\item Calculate the distance between the multivariate group means resulting 
from the linear model and store the result.\item For the desired number of permutations, default = 1000, randomize the residuals resulting from the linear model in step 3 and add those to the group means 
estimated in step 4.\item Calculate the difference in-group means and store the results.\item Use the resulting distribution as a null distribution for determining whether the two groups are further away in 
multivariate space than you would expect given the Var/Cov structure of the data.\item Repeat steps 3 -7 for each pair of groups generated in step 2.
\end{enumerate}

\code{PermuteLDA(Data, Groups, Nperm, Missing.Data="Complete")}
\begin{itemize}
\item \code{Data}  =  a (non-empty) numeric matrix of data values\item \code{Groups} = a (non-empty) vector indicating which group the rows in Data belong to\item \code{Nperm} = Number of permutations used 
to generate the null distribution \item \code{Missing.Data} = Missing data must either be removed or imputed.  Complete uses \code{CompleteData} to impute missing data, Remove censors individuals with 
missing values for one or more traits.
\end{itemize}

\section[Visualizations Implemented in multiDimBio]{Visualizations implemented in \pkg{multiDimBio}}
Visualization is an essential ingredient in any statistical analysis; however, multivariate data presents unique challenges to established visualization methods.  Here we have implemented a number of methods 
to serve either exploratory or illustrative purposes.  The visualizations rely heavily on the \CRANpkg{ggplot2} package (~\cite{ggplot2}).

\subsection{Exploratory Methods}
We have implemented three methods for exploring the data.  The first creates box and whisker plots of the data.  Importantly, the data can be either transformed using one of our pre-processing methods or left in 
raw form, see section~\ref{PP}.  We also provide methods for visualizing the loading of data onto principle components and discriminant functions.  

\paragraph[BoxWhisker]{\code{BoxWhisker}}
The \code{BoxWhisker} function will produce a box and whisker plot of each trait or variable in the data subdivided into groups, Figure~\ref{F1}.

\begin{figure}[tp]
\begin{center}
    \label{F1}\includegraphics[width=0.6\textwidth]{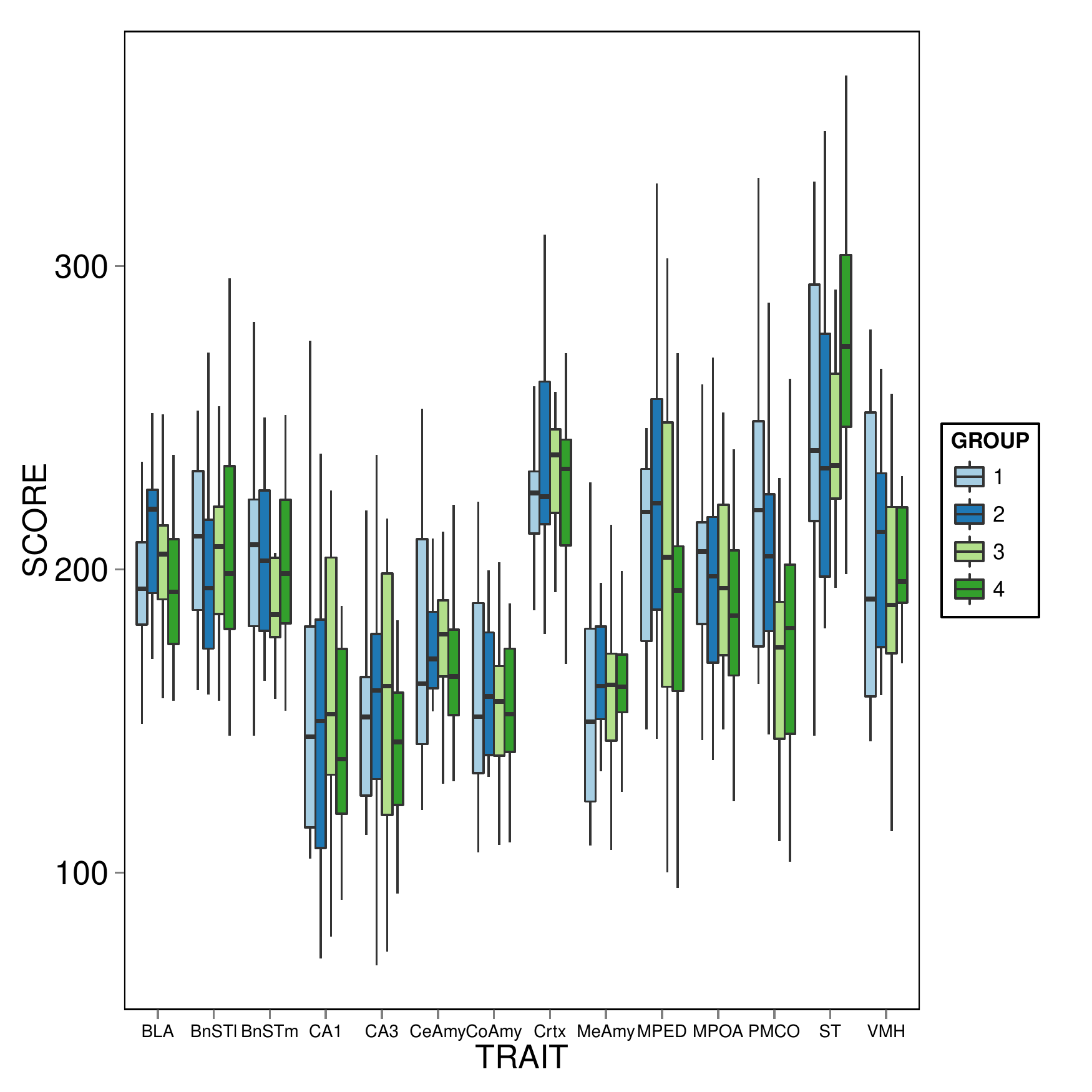}
\end{center}
\caption{
{\bf A box and whisker plot of brain nuclei activity.}  A box and whisker plot was created using the \code{BoxWhisker} function to illustrate the differences across brain regions and between groups.
}
\label{F1}
\end{figure}

\begin{example}
data("Nuclei")
data("Groups")
BoxWhisker(Nuclei, Groups, palette="Paired")
\end{example}

\paragraph[Loadings]{\code{Loadings}}
The \code{Loadings} function will produce a bar plot of each the loading of each trait or variable in the data onto the principle component or discriminant function axes.  The principle component analysis is 
performed using the \pkg{pcaMethods} package (~\cite{Stacklies2007},  Figure~\ref{F2}.

\begin{figure}[tp]
\begin{center}
    \label{F2}\includegraphics[width=1.0\textwidth]{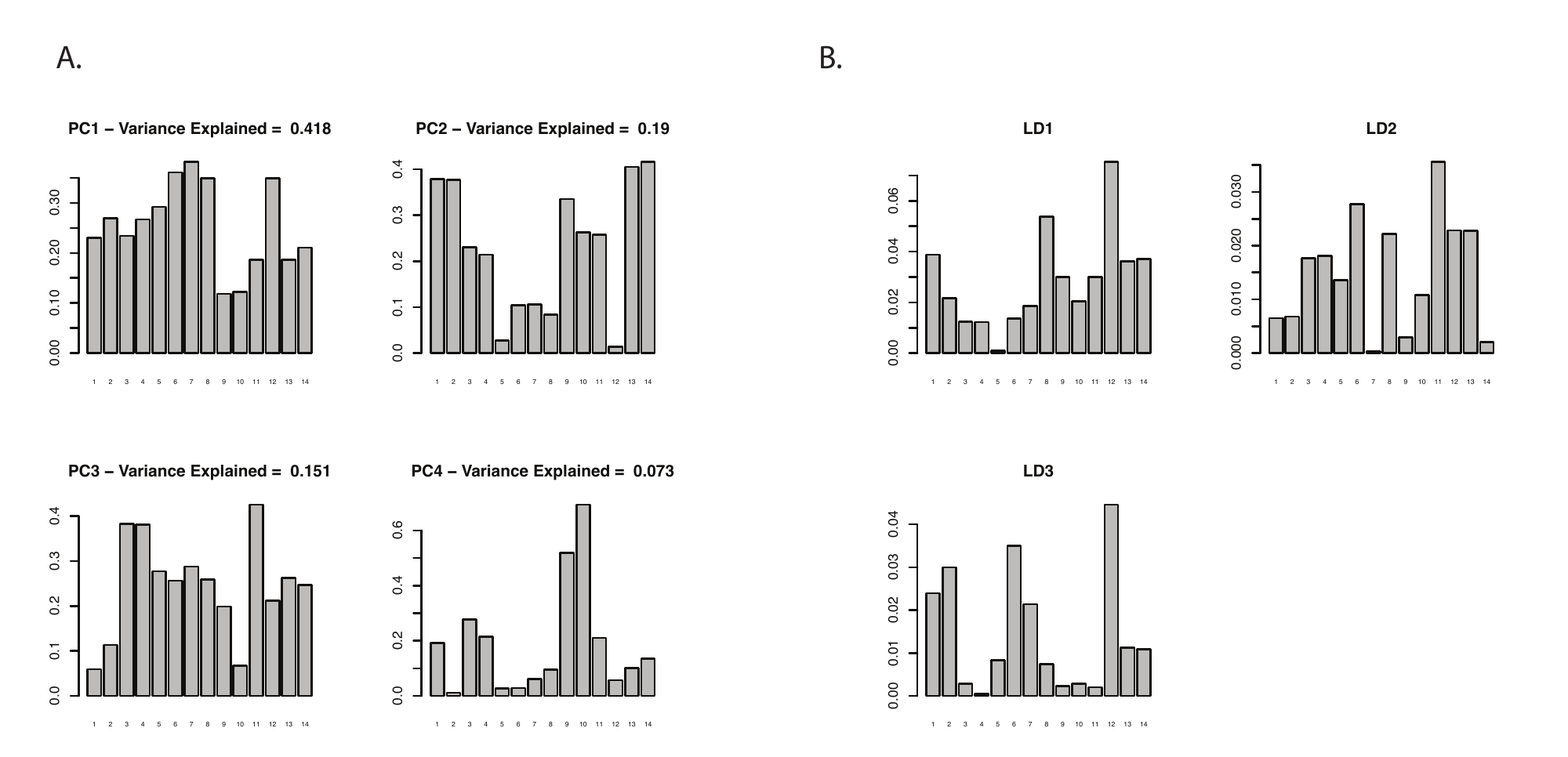}
\end{center}
\caption{
{\bf A bar plot of the loadings of brain nuclei activity on either PC or DA axes.}  A bar plot was created using the \code{barplot} function to illustrate the loading of each brain region onto the axes in either a 
principle component analysis, 2{\it a}, or discriminant analysis, 2{\it b}.
}
\label{F2}
\end{figure}

\begin{example}
data("Nuclei")
data("Groups")
Loadings(Nuclei, Groups, method=c("PCA", "LDA"))
\end{example}

\subsection{Illustrative Methods}

\paragraph{MANOVA visualizations}
To visualize the results of MANOVAs we use a series of interaction plots.  The interaction plots illustrate the effect of different treatments on the response variables.  This method is most effective when the 
response variables are first transformed into principle components; however, the method will work with any form of data.  

\code{IntPlot(Scores, Cov.A, Cov.B, pvalues=rep(1,8), int.pvalues=rep(1,4))}
\begin{itemize}
\item \code{Scores}  = Data for the analysis, preferably in the form of principle component scores\item \code{Cov.A} = Vector indicating the bivariate indicator for each row in Scores.\item \code{Cov.B} = Vector 
indicating the bivariate indicator for each row in Scores.\item \code{pvalues} = A vector of p values.  For each column in Scores there should be a p value for Cov.A and Cov.B. \item \code{int.pvalues} = A vector 
of p values for the interaction terms.  For each column in Scores there should be an entry in int.pvalues.
\end{itemize}

\begin{figure}[tp]
\begin{center}
   \label{F3}\includegraphics[width=1.0\textwidth]{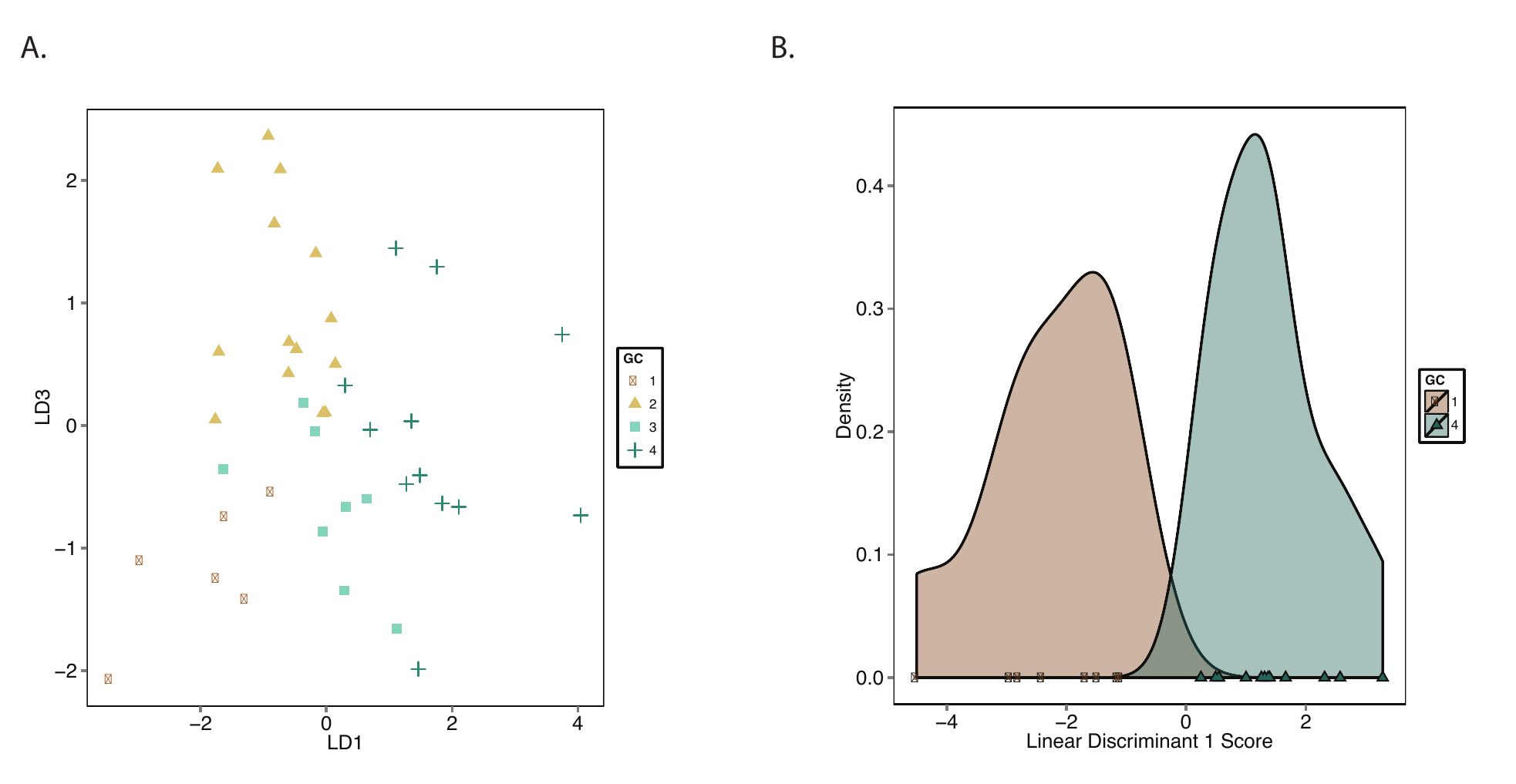}
\end{center}
\caption{
{\bf Two visualizations of a discriminant analysis.}  Plots were created using the \code{ldaPlot} function to illustrate how well groups can be distinguished using a discriminant analysis.  In panel {\it a}, all four 
groups are represented in a bivariate scatter plot, with LD1 scores on the x-axis and LD2 scores on the y-axis.  The comparisons can be set using arguments to \code{ldaPlot}.  To compare the separation of each 
pair of groups, a separate discriminant analysis is run for each unique combination of two groups.  The results are shown in panel {\it b}, where a density plot for each of the two groups can be compared to the 
actual observations of individuals in those groups, marked as points along the y=0 line. 
}
\label{F3}
\end{figure}

\paragraph{Discriminant function visualizations}
The visualization of discriminant analysis results is implemented in the \code{ldaPlot} function.  The function takes as input the traits and group IDs and will perform a discriminant function analysis and visualize 
the results.  For the pair-wise comparison of groups we use density histograms with points along the x-axis denoting the actual data, Figure~\ref{F3}. For multi-group comparisons we plot a bivariate scatter for all 
pairwise combinations of discriminant axes. The color of plotting symbols can be altered using the palette argument and the linear discriminant comparisons using the axes argument (with max axis $=$ number 
of groups - 1).

\begin{example}
data("Nuclei")
data("Groups")
ldaPlot(Nuclei, Groups, palette=``BrBG", axes=c(1,2,2,3,1,3))
\end{example}

\subsection{Landscape Plots} \label{LP}
To visualize the composite phenotype for each group, and thus compare the phenotype between groups, a functional landscape can be constructed. The peak of each node in the landscape is calculated as the 
percent of maximum from the highest group mean. The width of each node was adjusted to optimally fit the number of nodes in each landscape and has no statistical significance. A percent change landscape is 
then created to visualize the differences between groups. The direction of the node, either below or above the plane, indicates in which direction the mean is influenced by the effect of treatment or group. A node 
above the plane indicates the mean in the treated group is higher than the mean of the control group and vice versa. This method allows one to visualize the composite change in the phenotype of a control 
group to that of a treated group.  This graphical method is implemented in the \code{LandscapePlot} function, which relies on the \code{persp} function in the \CRANpkg{graphics} package to render the three-dimensional image.

\begin{example}
data("Nuclei")
data("Groups")
Data.Z <- ZTrans(Nuclei)
LandscapePlot(Data.Z[,1:6], Groups)
\end{example}

\section{Example: an analysis of neural response to stress}
We illustrate the methods and visualizations in the package using an experiment 
designed to disentangle the effects of ancestral (inherited) versus proximate (acquired) environmental stressors on the phenotype (~\cite{Crews2012}). The hypothesis is that ancestral exposures cause epigenetic 
reprogramming that changes how descendant individuals respond to life challenges. A two ``hit" paradigm was used. The first ``hit" consisted of embryonic exposure to a common-use 
fungicide Vinclozolin. This has been shown to create an epigenetic imprint that is incorporated into the germline and, hence, is manifest each generation in the absence of the 
original causative agent (~\cite{Skinner2010}). The second ``hit", 3 generations removed from the first, consists of chronic restraint stress (CRS) during adolescence. Stress in adolescence has powerful and 
permanent effects on brain and behavior, including epigenetic modifications to the nervous system (~\cite{Cicchetti2001}). 
We have provided four data sets from the above study to illustrate the methods in this paper.  Briefly, the data set \code{Nuclei} is a numeric matrix containing observed brain activity in 14 nuclei for 71 
individuals, the \code{Groups} object contains the group ID for each of the 71 individuals in \code{Nuclei}, \code{CondA} is a factor vector indicating whether and animal is from a lineage treated with the 
fungicide Vinclozolin, \code{CondB} is a factor vector indicating whether an animal was subjected to CRS, \code{Scores} are the results of a PPCA of \code{Nuclei}, and \code{Dyad} is a factor vector indicating 
housing dyad. The below example uses these data to explore the hypothesis that Vinclozolin and stress affect brain chemistry.  A detailed discussion of the data, methods, and results of this study can be found in ~\cite{Crews2012}.

Before embarking on any research, one of the first  statistical steps is to determine the necessary sample size for the desired experiments.  This can be accomplished using the \code{Power} function.  
Importantly, these calculations can be used in animal care and use protocols and for grant applications. 

\begin{example}
Power(func = "PermuteLDA", N = "DEFAULT.N", effect.size = 0.8, trials = 100)
Power(func = "FSelect", N = "DEFAULT.N", effect.size =0.8, trials = 100)
\end{example}

The results of the power analysis demonstrate differences between \code{FSelect} and \code{PermuteLDA} in the trade-off between sensitivy and specificity as the number of traits with significant effects increase, Figure~\ref{F5}.  Specifically,  \code{PermuteLDA} has faster rate of increase in power.  Importantly, the type-I error, or false positive, rate is very low for both methods, being effectively 0 for \code{FSelect} and much less than 10 percent for \code{PermuteLDA}, Figure~\ref{F5} black lines.  

Before pre-processing the data and performing statistical tests, it is important to visualize the data.  Our package provides methods for visualizing the raw and transformed data, Figure ~\ref{F1}.

\begin{figure}[tp]
\begin{center}
    \label{F5}\includegraphics[width=1.0\textwidth]{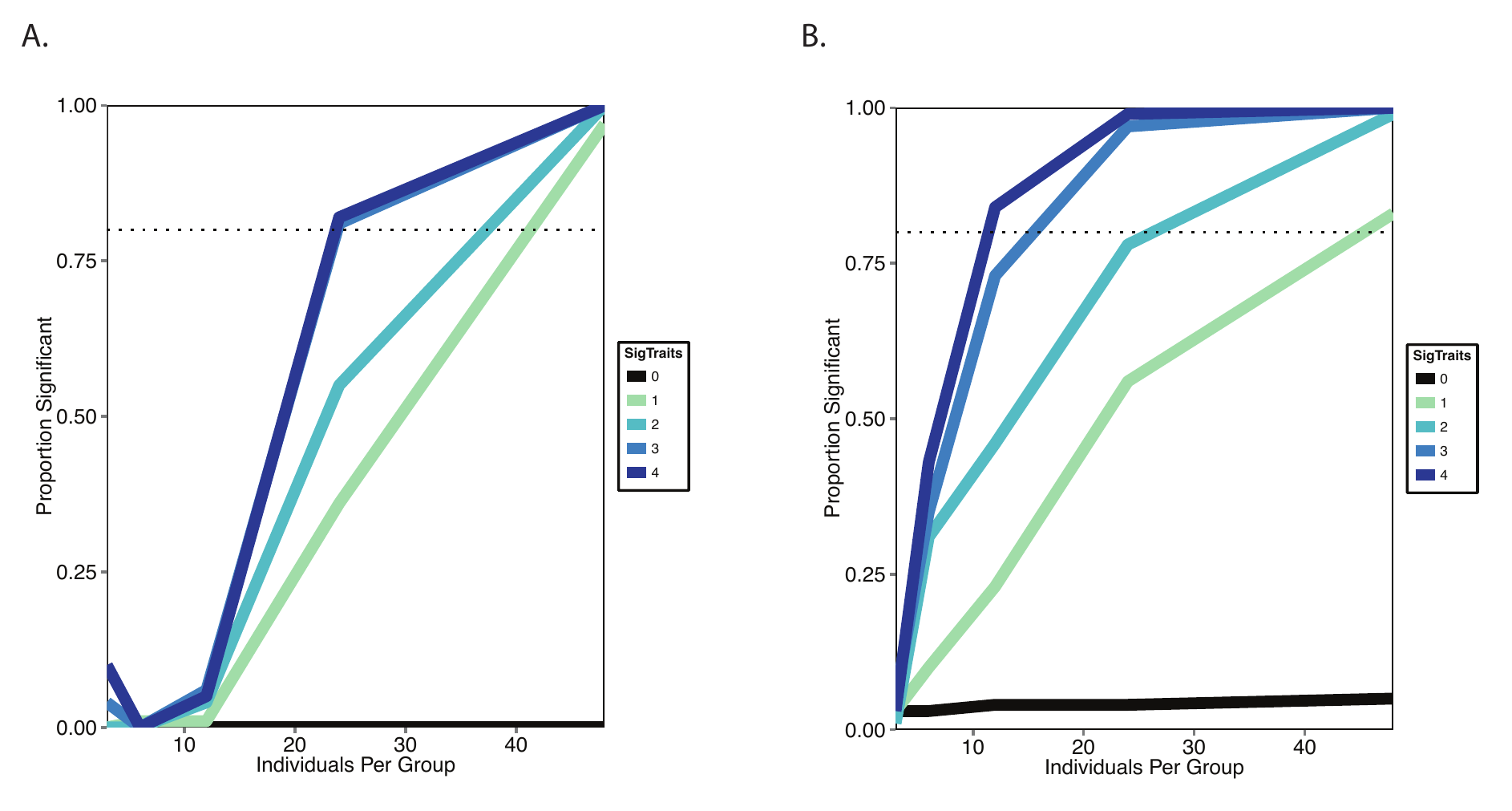}
\end{center}
\caption{
{\bf Power analysis of \code{FSelect} and \code{PermuteLDA}.}  A power analysis was performed for, 5{\it a}, \code{FSelect} and, 5{\it b} \code{PermuteLDA}.  The effect size was 0.8 with the number of 
individuals per group ranging between  6 and 96.  The line color indicates how many out of six traits were significantly different between the groups.  The black line is for zero significant traits and is a measure 
of the type-I error rates.  The green and blue lines are for 1 - 4 significant traits and are measures of the type-II error rate.  The 80 percent power level is indicated by a dashed, horizontal line in each figure.
}
\label{F5}
\end{figure}

\begin{example}
data("Nuclei")
data("Groups")
BoxWhisker(Nuclei,Groups)
\end{example}

To impute missing data use the function \code{CompleteData}.  The output will be a figure illustrating the performance of the method and matrix of complete observations, Figure~\ref{F6}.  This function relies on 
methods implemented in the \pkg{MASS} and \pkg{pcaMethods} packages (~\cite{MASS, Stacklies2007}).

\code{CompleteData(Nuclei,cut.trait=0.5, cut.ind=0.5, NPCS=4, show.test=TRUE)}

Here we  present an example where all the tests included in this package are performed successively on data from the aforementioned experiment.  

\begin{example}

data(Nuclei)
data(Groups)
DAT.comp<-CompleteData(Nuclei, NPCS=4)
Groups.use<-c(1,2)
use.DAT<-which(Groups==Groups.use[1]|Groups==Groups.use[2])
DAT.use<-Nuclei[use.DAT,]
GR.use<-Groups[use.DAT]
FSelect(DAT.use,GR.use,3)

Missing data imputed using CompleteData to exlude missing data 
set Missing.Data=Remove 
 
Final lda model saved as an R object 
 
An object of class "FSELECT"
Slot "Selected":
[1]  7  6 11

Slot "F.Selected":
[1] 1.2137438 0.3684986 0.1744205

Slot "PrF":
[1] 1 1 1

Slot "PrNotes":
[1] "PrF has been holm adjusted for 3 comparisons"

\end{example}

\begin{figure}[tp]
\begin{center}
    \label{F6}\includegraphics[width=0.8\textwidth]{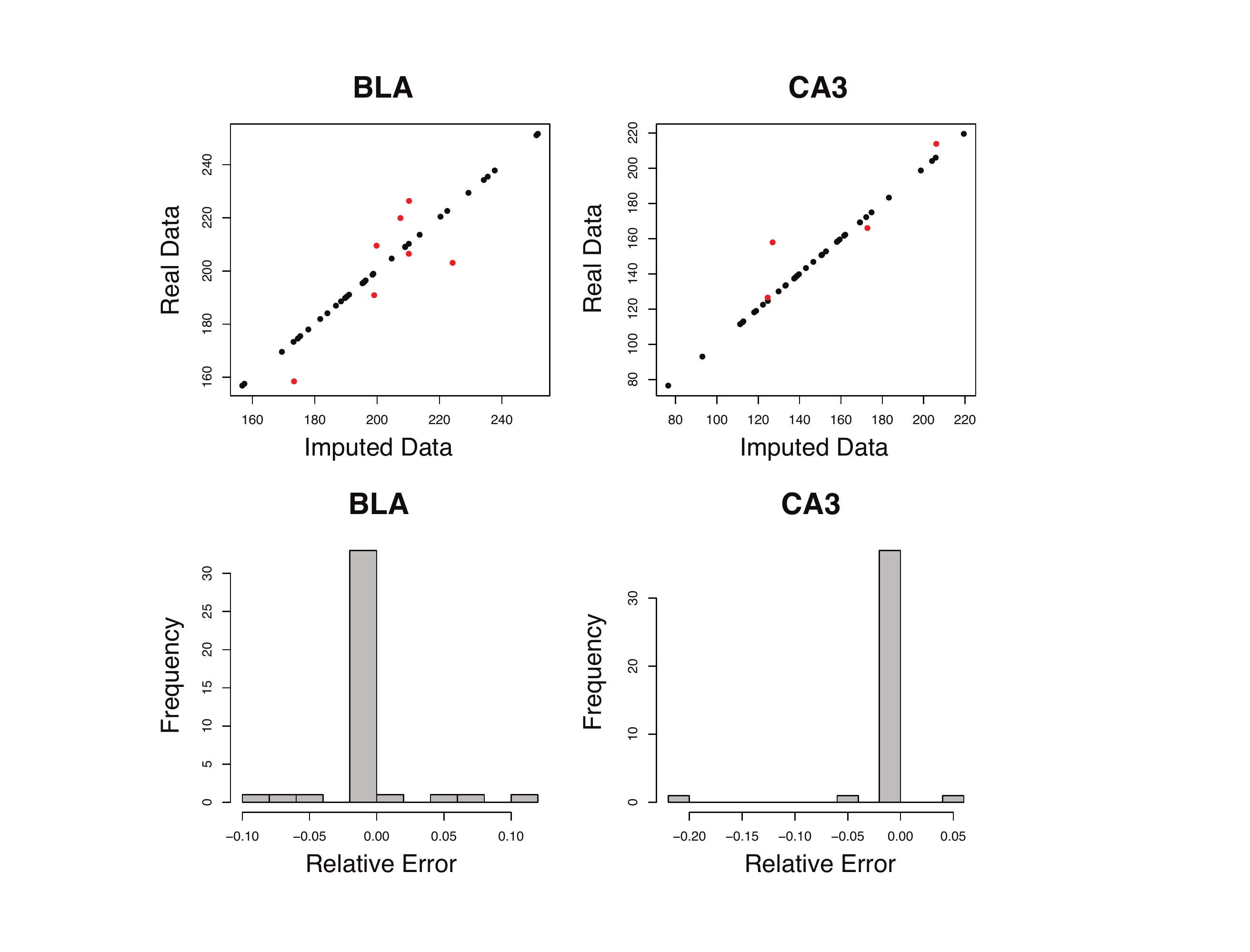}
\end{center}
\caption{
{\bf The results of imputing missing data using \code{CompleteData}.}  The \code{CompleteData} function imputes missing data using a probabilistic principle component  framework.   Using a complete data set, 
in this case the \code{Nuclei} data with all individuals with missing values removed, it is possible to assess the performance of the method.  For two traits, BLA and CA3, the observed vs. expected are plotted for 
actual data, black, and imputed data, red, histograms of observed - expected are also provided.  Although only two of the 14 nuclei are shown here, the method outputs a .pdf file for each trait, in this case 14 
separate graphs would be created.
}
\label{F6}
\end{figure}

\begin{example}

PermuteLDA(Nuclei,Groups,100)

   Group 1 Group 2         Pr Distance
1       1       2 0.89108911 30.85968
2       1       3 0.91089109 32.27643
3       1       4 0.22772277 53.90860
4       2       3 0.57425743 42.87241
5       2       4 0.03960396 66.60140
6       3       4 0.50495050 46.34542

\end{example}

\begin{figure}[tp]
\begin{center}
    \label{F7}\includegraphics[width=0.8\textwidth]{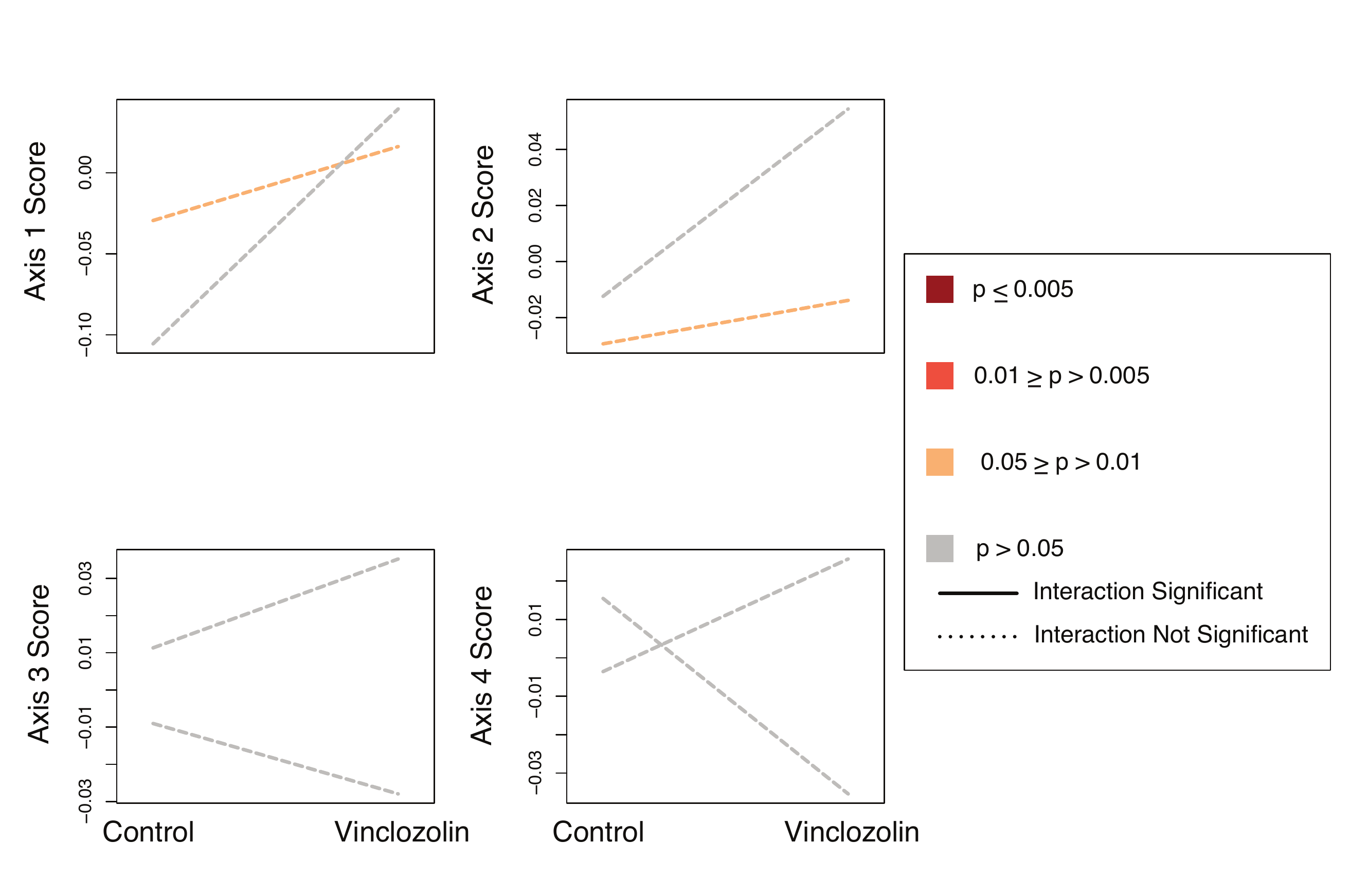}
\end{center}
\caption{
{\bf MANOVA results presented as a series of interaction plots.}  MANOVA results are presented for four principle component axes.  The lines indicate stress response across Control and Vinclozolin lines, with a 
putative interaction being indicated by intersecting lines.  The color of the line indicates p-values and the line type indicates the interaction significance, dashed = non-significant and solid = significant.  The first 
two principle components exhibit a significant difference in neural response to stress in Vinclozolin vs Control lines.  Although none of the interactions were significant, there was a trend towards significance for 
axis four, this is evidenced by the crossing lines in the lower right panel.
}
\label{F7}
\end{figure}

\begin{example}

data("CondA")
data("CondB")
data("Scores")
data("Dyad")
m1<-manova(Scores~CondA*CondB+Dyad)
summary(m1)

            Df  Pillai approx F num Df den Df    Pr(>F)    
CondA        1 0.38916   4.7781      4     30  0.004206 ** 
CondB        1 0.18402   1.6914      4     30  0.178001    
Dyad        34 3.15420   3.6196    136    132 3.628e-13 ***
CondA:CondB  1 0.28280   2.9573      4     30  0.035818 *  
Residuals   33                                             
---
Signif. codes:  0 ''***'' 0.001 ''**'' 0.01 ''*'' 0.05 ''.'' 0.1 '' '' 1 

\end{example}

The final step is to visualize the results.  The approach is a combination of interaction plots for MANOVA results and either density or scatter plots for the discriminant function analysis.  The results for the
discriminant function analysis can be seen in Figure~\ref{F3}.  The interaction plots for the MANOVA results can be seen in Figure~\ref{F7}.  The p-values for the different covariates must be given as arguments 
to the function \code{IntPlot}.  To illustrate the effect of Vinclozolin on stress response, the only significant difference identified using \code{PermuteLDA}, we created a functional landscape (see section~\ref{LP}) 
using the \code{LandscapePlot} function, see Figure~\ref{F8}.  This landscape uses the six brain nuclei that best distinguish these two groups, as determined using the \code{FSelect} function.

\begin{example}

data("Scores")
data("CondA")
data("CondB")
IntPlot(Scores,CondA,CondB,pvalues=c(0.03,0.6,0.05,0.07,0.9,0.2,0.5,0.3),
 int.pvalues=c(0.3,0.45,0.5,0.12))

data("Nuclei")
data("Groups")
Data.Z<-ZTrans(Nuclei)
Traits<-c(1,2,9,10,11,13)
LandscapePlot(Data.Z[,Traits], Groups)

\end{example}

\section{Discussion}

\begin{figure}[tp]
\begin{center}
    \label{F8}\includegraphics[width=0.8\textwidth]{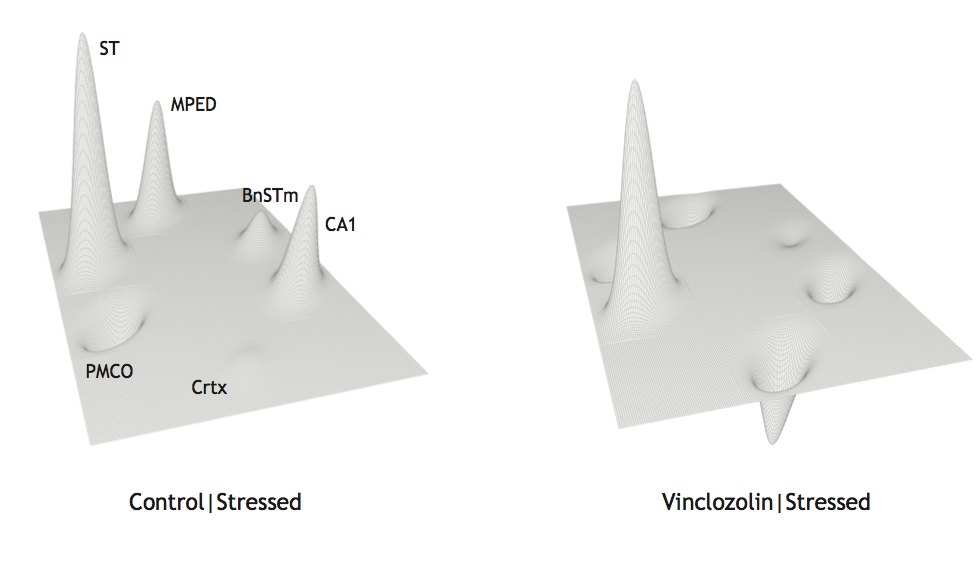}
\end{center}
\caption{
{\bf Landscape Plot of Brain Nuclei}  A three-dimensional landscape plot, called a Functional Landscape, was created to illustrate the effect Vinclozolin on stress response, the only significant difference 
identified using \code{PermuteLDA}.  The six most important brain regions were: MPED, ST, BnSTm, PMCO, CA1, and Ctrx.  These regions were identified using the \code{FSelect} function.  The raw data were 
transformed into Z-scores before creating the landscape plots.
}
\label{F8}
\end{figure}

Our package is a tool for researchers interested in applying a systems approach to their research. We provide methodology for multivariate analysis, visualization, and data pre-processing. Our quantitative 
framework supports a research project from inception through publication, suggesting visualizations and statistical methods along the way.  With the growing size of data sets collected under a range of 
scientific experiments, the emerging challenge is how to analyze and visualize that data.  Here we are not simply referring to processing large quantities of data, but are instead referring to the challenge of 
distinguishing patterns from noise.  Presumably the experiment will not affect all traits measured and those showing a response could do so in different directions and/or with different magnitudes.  Therefore, we advocate for a dimension reduction approach that seeks to find patterns in the aggregate while simultaneously removing those variables which either fail to improve the model fit or are misleading.

Visualizations are an important, but often overlooked, aspect of communicating the results of scientific 
research.   A great visualization should be informative and functional; informative in that it effectively conveys the results and functional in that it uses our understanding of human behavioral psychology to 
engage the reader.  The image must also use colors and symbols that are evident to a broad audience.  We have provided some suggested visualizations for the types of multivariate data common in biological 
studies.  Important future development of this package will incorporate additional graphical methods.  

One important caveat is that our methods are somewhat lacking in statistical power.  For effect sizes typical of behavioral studies, $\sim$35 individuals per group 
would be needed for \code{FSelect} and 20 individuals per group for \code{permuteLDA} to achieve 80 percent power, see Figure ~\ref{F5}.  However, our methods perform very well with respect to minimizing 
the type-I error rate.  Future  work should focus on adapting existing methods and developing new methods to allow for smaller sample and effect sizes.  As stated earlier this package is meant to be a project that 
will grow and develop; therefore, we welcome suggestions and contributions and plan regular updates to both the statistical and visual methods.  

\bibliography{scarpino-gillete-crews.bib}

%% file: RJwrapper.bbl
\begin{thebibliography}{24}
\providecommand{\natexlab}[1]{#1}
\providecommand{\url}[1]{\texttt{#1}}
\expandafter\ifx\csname urlstyle\endcsname\relax
  \providecommand{\doi}[1]{doi: #1}\else
  \providecommand{\doi}{doi: \begingroup \urlstyle{rm}\Url}\fi

\bibitem[Cicchetti and Walker(2001)]{Cicchetti2001}
D.~Cicchetti and E.~Walker.
\newblock Editorial: Stress and development: Biological and psychological
  consequences.
\newblock \emph{Development and Psychopathology}, 13:\penalty0 413--418, 2001.

\bibitem[Collyer and Adams(2007)]{CollyerAdams2007}
M.~Collyer and D.~Adams.
\newblock Analysis of two-state multivariate phenotypic change in ecological
  studies.
\newblock \emph{Ecology}, 88\penalty0 (3):\penalty0 683--692, 2007.

\bibitem[Costanza and Afifi(1979)]{Costanza1979}
M.~Costanza and A.~Afifi.
\newblock Comparison of stopping rules in forward stepwise discriminant
  analysis.
\newblock \emph{Journal of the American Statistical Association}, pages
  777--785, 1979.

\bibitem[Crews and McLachlan(2006)]{Crews2006}
D.~Crews and J.~McLachlan.
\newblock Epigenetics, evolution, endocrine disruption, health, and disease.
\newblock \emph{Endocrinology}, 147\penalty0 (6):\penalty0 s4--s10, 2006.

\bibitem[Crews et~al.(2006)Crews, Lou, Fleming, and Ogawa]{Crews2006b}
D.~Crews, W.~Lou, A.~Fleming, and S.~Ogawa.
\newblock From gene networks underlying sex determination and gonadal
  differentiation to the development of neural networks regulating sociosexual
  behavior.
\newblock \emph{Brain Research}, pages 109--121, 2006.

\bibitem[Crews et~al.(2012)Crews, Gillette, Scarpino, Manikkam, Savenkova, and
  Skinner]{Crews2012}
D.~Crews, R.~Gillette, S.~Scarpino, M.~Manikkam, M.~Savenkova, and M.~Skinner.
\newblock Epigenetic transgenerational alterations to stress response in brain
  gene networks and behavior.
\newblock \emph{Proc. Natl. Acad. Sci. USA}, 109\penalty0 (23):\penalty0
  9143--9148, 2012.

\bibitem[Everitt and Hothorn(2011)]{Everitt2011}
B.~Everitt and T.~Hothorn.
\newblock \emph{An Introduction to Applied Multivariate Analysis with R (Use
  R)}.
\newblock Springer, 2011.

\bibitem[Gottesman(1997)]{Gottesman1997}
I.~Gottesman.
\newblock Twins: En route to qtls for cognition.
\newblock \emph{Science}, 276:\penalty0 1522--1523, 1997.

\bibitem[Grossman et~al.(2003)Grossman, Churchill, McKinney, Kodish, Otte, and
  Greenough]{Grossman2003}
A.~Grossman, J.~Churchill, B.~McKinney, I.~Kodish, S.~Otte, and W.~Greenough.
\newblock Experience effects on brain development: Possible contributions to
  psychopathology.
\newblock \emph{J. Child Psych.}, 44:\penalty0 33--63, 2003.

\bibitem[Habbema and Hermans(1977)]{Habbema1977}
J.~Habbema and J.~Hermans.
\newblock Selection of variables in discriminant analysis by f-statistics and
  error rate.
\newblock \emph{Technometrics}, 19\penalty0 (4):\penalty0 487--493, 1977.

\bibitem[Jennrich(1977)]{Jennrich1977}
R.~Jennrich.
\newblock \emph{Stepwise Discriminant Analysis}, volume~3.
\newblock New York: John Wiley \& Sons, 1977.

\bibitem[Kitano(2002)]{Kitano2002}
H.~Kitano.
\newblock Systems biology: A brief overview.
\newblock \emph{Science}, 1\penalty0 (295):\penalty0 1662--1664, 2002.

\bibitem[Moore et~al.(1997)Moore, Brodie, and Wolf]{Moore1997}
A.~Moore, E.~Brodie, and J.~Wolf.
\newblock Interacting phenotypes and the evolutionary process: I. direct and
  indirect genetic effects of social interactions.
\newblock \emph{Evolution}, 51\penalty0 (5):\penalty0 1352--1362, 1997.

\bibitem[Nijhout(2003)]{Nijhout2003}
H.~Nijhout.
\newblock The importance of context in genetics.
\newblock \emph{American Scientist}, 91:\penalty0 416--418, 2003.

\bibitem[Nijhout(2005)]{Nijhout2005}
H.~Nijhout.
\newblock Problems and paradigms: Metaphors and the role of genes in
  development.
\newblock \emph{BioEssays}, 12\penalty0 (9):\penalty0 441--446, 2005.

\bibitem[Roweis(1997)]{Roweis1997}
S.~Roweis.
\newblock Em algorithms for pca and sensible pca.
\newblock \emph{Neural Inf. Proc. Syst.}, 10:\penalty0 626--632, 1997.

\bibitem[Skinner et~al.(2010)Skinner, Manikkam, and
  Guerrero-Boasagna]{Skinner2010}
M.~Skinner, M.~Manikkam, and C.~Guerrero-Boasagna.
\newblock Epigenetic transgenerational actions of environmental factors in
  disease etiology.
\newblock \emph{Trends in Endocrinology and Metabolism}, 21\penalty0
  (4):\penalty0 214--222, 2010.

\bibitem[Stacklies et~al.(2007)Stacklies, Redestig, Scholz, Walther, and
  Selbig]{Stacklies2007}
W.~Stacklies, H.~Redestig, M.~Scholz, D.~Walther, and J.~Selbig.
\newblock \pkg{pcaMethods} -- a bioconductor package providing pca methods for
  incomplete data.
\newblock \emph{Bioinformatics}, 23:\penalty0 1164--1167, 2007.

\bibitem[Team(2011)]{stats}
R.~D.~C. Team.
\newblock \emph{R: A Language and Environment for Statistical Computing}.
\newblock R Foundation for Statistical Computing, Vienna, Austria, 2011.

\bibitem[Tippping and Bishop(1999)]{Tipping1999}
M.~Tippping and C.~Bishop.
\newblock Probabilistic principle componenet analysis.
\newblock \emph{Journal of the Royal Statistical Society B}, 61\penalty0
  (3):\penalty0 611--622, 1999.

\bibitem[Troyanskaya et~al.(2001)Troyanskaya, Cantor, Sherlock, Brown, Hastie,
  Tibshirani, Botstein, and Altman]{Troyanskaya2001}
O.~Troyanskaya, M.~Cantor, G.~Sherlock, P.~Brown, T.~Hastie, R.~Tibshirani,
  D.~Botstein, and R.~Altman.
\newblock Missing value estimation methods for dna microarrays.
\newblock \emph{Bioinformatics}, 17\penalty0 (6):\penalty0 520--5252, 2001.

\bibitem[Venables and Ripley(2002)]{MASS}
W.~Venables and B.~Ripley.
\newblock \emph{Modern Applied Statistics with S}.
\newblock Springer, fourth edition, 2002.

\bibitem[Waddington(1957)]{Waddington1957}
C.~Waddington.
\newblock \emph{The Strategy of the Genes: A Discussion of Some Aspects of
  Theortical Biology}.
\newblock London: Allen and Unwin, 1957.

\bibitem[Wickham(2009)]{ggplot2}
H.~Wickham.
\newblock \emph{\pkg{ggplot2}: Elegant Graphics for Data Analysis}.
\newblock Springer, 2009.

\end{thebibliography}
